# Tuning the pressure-induced superconductivity in Pd-substituted $CeRhIn_5$

*M. Kratochvilova[1], K. Uhlirova[1], J. Prchal[1], J. Prokleska[1], Martin Misek[1], V. Sechovsky[1]*

[1] *Department of Condensed Matter Physics, Faculty of Mathematics and Physics, Charles University in Prague, Ke Karlovu 5, 121 16 Prague 2, Czech Republic*

**Keywords:** heavy fermion superconductivity; antiferromagnetism, pressure–temperature phase diagram, $CeRhIn_5$

**Abstract**

The effect of substituting Rh in $CeRh_{1-x}Pd_xIn_5$ with Pd up to $x = 0.25$ has been studied on single crystals. The crystals have been grown by means of the In self-flux method and characterized by x-ray diffraction and microprobe. The tetragonal $HoCoGa_5$-type of structure and the $c/a$ ratio of the parent compound remains intact by the Pd substitution; the unit cell volume increases by 0.6 % with $x = 0.25$ of Pd. The low-temperature behavior of resistivity was studied also under hydrostatic pressure up to 2.25 GPa. The Pd substitution for Rh affects the magnetic behavior and the maximum value of the superconducting transition temperature measured at pressures above 2 GPa only negligibly. On the other hand, the results provide evidence that superconductivity in $CeRh_{0.75}Pd_{0.25}In_5$ is induced at significantly lower pressures, i.e. the Pd substitution for Rh shifts the $CeRh_{1-x}Pd_xIn_5$ system closer to coexistence of magnetism and superconductivity at ambient pressure.

## 1. Introduction

Due to tunable interplay between magnetism and superconductivity, the $Ce_nT_mIn_{3n+2m}$ ($n = 1, 2$; $m = 1$; $T =$ Co, Rh, Ir) heavy fermion (HF) materials have been attractive for thorough studies of varieties of the two cooperative phenomena. The compounds are known to be on the verge of a magnetic to a non-magnetic quantum critical point (QCP) [1] where an unconventional superconducting (SC) state has been reported. They crystallize in the tetragonal $Ho_nCo_mGa_{3n+2m}$-type of structure that can be viewed as a quasi-2D structure of $m$ layers of $TIn_2$ alternating with $n$ layers of $CeIn_3$ along the $c$-axis. Recently, the family of $Ce_nT_mIn_{3n+2m}$ compounds has turned out to be unexpectedly rich in new members and structure types like the lately discovered materials $Ce_3PtIn_{11}$, $Ce_3PdIn_{11}$ and $Ce_5Pd_2In_{19}$ which show rather complicated layer stackings [2, 3]. $CeRhIn_5$ belongs, on the other hand, to the well-known prototypical HF systems within the $Ce_nT_mIn_{3n+2m}$ group. It orders antiferromagnetically (AF) below $T_N = 3.8$ K [1] and exhibits a complex magnetic structure when the magnetic field is applied along the $a$-axis [4, 5]. The ground state of $CeRhIn_5$ is incommensurate (½, ½, 0.297) antiferromagnetic with the magnetic moments oriented helicoidally in the basal plane [4]. In higher magnetic fields, $T_N$ increases and two magnetic-field-induced transitions appear at $T_1$ and $T_2$. The phase transitions divide the phase diagram into several regions with different magnetic structures. The magnetic order changes from helicoidal to elliptical between $T_N$ and $T_2$. A commensurate antiferromagnetic structure with $q = $ (½, ½, ¼) emerges in the high-field phase below $T_1$. In contrast, the response of the system to a magnetic field applied parallel to the $c$-axis is modest, decreasing $T_N$ monotonically. The frustration of nearest-neighbor and next-nearest-neighbor exchange interactions along the *c*-axis likely stands behind this magnetic structure complexity as was recently confirmed via high-resolution neutron spectroscopy measurements [6]. The obtained spin wave spectra can be quantitatively reproduced with a simple $J_1$-$J_2$ model that explains the magnetic spin-spiral magnetic ground state of CeRhIn$_5$ along *c* and enables to determine all microscopic RKKY exchange interactions.

A broad variety of substitution experiments [7-17] has been performed on CeRhIn$_5$. Substitution of Ir for Rh induces two different SC states. In the Rh-rich region, the superconductivity is induced by applied hydrostatic pressure while in the Ir-rich region a SC state has been observed at ambient pressure, being sensitive to applied pressure [7]. Co substitution leads to monotonic suppression of magnetism and development of superconductivity with a broad concentration range in which both phenomena coexist [8]. Substitutions at the transition-metal site affect the magnetism rather moderately but, in contrast, very low concentrations of substitution at the rare-earth or the In site are sufficient to change the transition temperatures strongly [7-9]: La substitution changes the long-range magnetic order into coexistence of short-range order and non-Fermi liquid behavior (nFl) [10]. Non-isoelectronic (hole) substitution of CeRhIn$_5$ with either Hg or Cd at the In site has a similar effect of non-monotonic variation of $T_N(x)$ which may be connected with an incommensurate-to-commensurate magnetic-order transition [11-13] induced by a magnetic field applied in the basal plane [18]. Electron substitution with Sn leads to suppression of magnetism [14] which is likely due to an increase of the Kondo coupling [9]. Recent studies of CeRhIn$_{5-x}$Sn$_x$ have revealed a change from an incommensurate to a commensurate magnetic structure in the vicinity of the magnetic QCP [15].

Besides substitution, the ground state of CeRhIn$_5$ can also be tuned by hydrostatic pressure [7, 19-21]. Application of pressure induces a SC phase which coexists with the AF phase up to a pressure of about 2 GPa. At higher pressures, only the superconductivity remains [21].

In the present work, we report on the interplay between superconductivity and antiferromagnetism in CeRh$_{1-x}$Pd$_x$In$_5$, extending the substitution scenario of the Ce*T*In$_5$ series (*T* = Co, Rh, Ir) to non-isoelectronic substitution at the transition-metal site. A combined study of the effects of hydrostatic pressure and non-isoelectronic Pd substitution on the low-temperature properties of CeRhIn$_5$ has been performed. CeRh$_{1-x}$Pd$_x$In$_5$ single crystals with *x* = 0, 0.1 and 0.25 have been prepared and studied by means of magnetization, electrical resistivity and specific heat as a function of temperature and applied magnetic field. The sample with *x* = 0.25 was also subjected to measurements of the electrical resistivity under hydrostatic pressure. Our results show that the superconducting transition temperature $T_c$ of CeRh$_{1-x}$Pd$_x$In$_5$ with *x* = 0.25 remains much higher for pressures below the critical pressure compared to *x* = 0. The effect of Pd substitution will be discussed in context of the presence of a textured SC phase below critical pressure $p_{c1}$ determined from the anisotropy of the measured in-plane and out-of-plane electrical resistivity of CeRhIn$_5$[22].

## 2. Sample synthesis and characterization

Similar to other Ce*T*$_{1-x}$*T'*$_x$In$_5$ (*T*, *T'* = Co, Rh, Ir) compounds, single crystals of CeRh$_{1-x}$Pd$_x$In$_5$ were grown by means of the In self-flux method. While the growth of Ce*T*$_{1-x}$*T'*$_x$In$_5$ is rather straightforward [7], the growth of CeRh$_{1-x}$Pd$_x$In$_5$ is more challenging because of some difficulties that have been tackled earlier in the synthesis of the $Ce_nPdIn_{3n+2}$ ($n = 2, 3$) materials [3, 23]. The initial molar ratio of $Ce:Rh_{1-y}Pd_y:In$ was 1:1:30-40 ($y = 0.1-1$). The batches were heated up to 950°C, kept at this temperature for 10 hours to let the mixture homogenize properly and then cooled down slowly (~3°C/h) to 600°C, where the remaining flux was decanted. The structure type and lattice parameters of the single crystals were determined by single crystal x-ray diffraction using Rigaku RAPID II. Microprobe analysis was performed using a Scanning Electron Microscope (SEM) Tescan Mira LMH equipped with energy dispersive x-ray detector (EDX) Bruker AXS and ESPRIT software. The samples selected for further measurements were checked carefully in order to determine the actual Pd content and to eliminate the ones with traces of spurious phases. The actual Pd content in the samples has been found to be lower than the nominal Pd concentration of the melt (similar to other substitutions [11-13]). In total, three single crystals, $CeRhIn_5$ and two samples with Pd contents $x = (10 \pm 2)\%$ and $x = (25 \pm 5)\%$, have been selected for further measurements. The samples with $x > 0.25$ contained inclusions of $Ce_2Rh_{1-x}Pd_xIn_8$ and $CeIn_3$ in such a form that they could not be isolated. We could find samples with the maximum Pd content $x \sim 0.5$ detected by microprobe analysis; however, these samples were not measured because of the inclusions.

The attempts to grow Pd-rich single crystals have been unsuccessful so far. The characteristic x-ray lines in Rh and Pd EDX spectra partially overlap ($L_{\alpha1} = 2.838$ keV for Pd and $L_{\beta1} = 2.834$ keV for Rh, as shown in figure 1), therefore the measurement uncertainty of the quantitative microprobe analysis can be significant. To verify the analysis given by the ESPRIT software, we prepared Rh-Pd binary alloys with known compositions $Rh_{0.9}Pd_{0.1}$ and $Rh_{0.75}Pd_{0.25}$ and used them as standards. The error of determining the Pd content by the ESPRIT software was taken into account when determining the Pd amount in the substituted samples.

The spectra of Rh and Pd L-series obtained from Rh, Pd, their alloys $Rh_{0.75}Pd_{0.25}$ and $Rh_{0.9}Pd_{0.1}$, and from $CeRh_{1-x}Pd_xIn_5$ single crystals are presented in figure 1.

Noticeably, although the Rh and Pd lines overlap in the spectra, the increase of Pd content on the expense of Rh is evident: While the intensity of the Rh $L_{\alpha1}$ line decreases simultaneously, the bump at Pd $L_{\beta1}$ line (2.990 keV) increases.

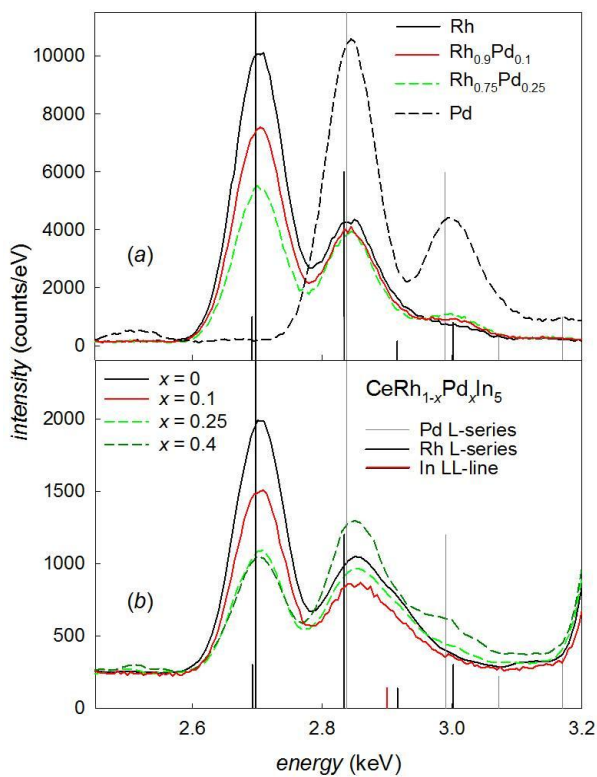

**Figure 1.** (*a*) The EDX point spectra of polycrystalline standards $Rh_{0.9}Pd_{0.1}$, $Rh_{0.75}Pd_{0.25}$ and of pure Rh and Pd metal. (*b*) The EDX point spectra of $CeRh_{1-x}Pd_xIn_5$ ($x = 0, 0.1, 0.25, 0.4$) single crystals. The vertical lines mark the energies and intensities of the characteristic x-ray lines (from the largest to the smallest intensity: $L_{\alpha1}+L_{\alpha2}$, $L_{\beta1}$, $L_{\beta2}$, $L_{\beta3}$ for Pd (Rh); $L_l$ for In).

The peak at Pd $L_{\alpha1}$ line in the single crystal spectra is broadened due to the Indium $L_l$-line (see figure 1(b)) which plays a non-negligible role considering the large stoichiometric amount of indium in the compound.

The magnetization was measured in the temperature range 1.8-300 K using a MPMS 7 T (*Quantum Design*). Measurements of the electrical resistivity and specific heat at ambient pressure were performed in the temperature range 0.35-300 K in a PPMS 9 T apparatus equipped with low-temperature $^3$He option. The electrical resistivity was measured by means of a conventional four-probe AC method at both ambient and hydrostatic pressure.

For hydrostatic pressure experiments down to 1.8 K, static pressures *p* up to 2.25 GPa were generated using a hybrid two-layered Cu-Be/NiCrAl clamped pressure cell [24] with a highest nominal pressure of 3 GPa, which has been designed to fit into the PPMS 9T. Daphne oil 7373 was used as a pressure-transmitting medium. The difference between the room- and the lowest measured temperature of the Daphne 7373 oil inside the pressure cell is $\Delta p \sim 0.2$ GPa and remains similar all over the used range of pressures [25]. The pressure values in the low-temperature region were determined using the linear pressure dependence of the electrical resistivity of manganin at room temperature and the difference $\Delta p = 0.2$. The observed very narrow SC transitions indicate a good hydrostatic environment in the experiment.

## 3. Results and discussion

The x-ray diffraction results confirmed that the CeRh$_{1-x}$Pd$_x$In$_5$ ($x = 0.1, 0.25$) compounds adopt the tetragonal HoCoGa$_5$-type structure similar to CeRhIn$_5$. Both the lattice parameters and the unit cell volume slightly increase with increasing Pd content (see table 1). The effect of Pd substitution on the $c/a$ ratio is negligible contrary to Co- and Ir-substituted samples up to $x = 0.25$ [16, 17]. Such small change of the lattice parameters was observed also in related systems CeRh$_{1-x}$T$_x$In$_5$ ($T$ = Co, Ir) [7, 17] and in CeCoIn$_5$ substituted by Ru [26].

**Table 1.** Lattice parameters of CeRh$_{1-x}$Pd$_x$In$_5$.

| $x$ | $a$ (Å) | $c$ (Å) | $V$ (Å$^3$) |
|---|---|---|---|
| 0 | 4.652 | 7.542 | 163.2 |
| 0.1 | 4.653 | 7.544 | 163.3 |
| 0.25 | 4.663 | 7.553 | 164.2 |

As a first step, a series of ambient-pressure measurements has been performed on the single crystals with concentrations $x$ = 0, 0.1 and 0.25. The specific heat of CeRh$_{1-x}$Pd$_x$In$_5$ presented in figures 2 and 3(*a*) reveals a second-order AF transition. The idealization of the specific-heat jump under the constraint of entropy conservation yields an ordering temperature $T_N$ = 3.78 K for $x$ = 0 in zero magnetic field. The Néel temperature $T_N$ slightly decreases with increasing Pd content to 3.76 K for $x$ = 0.1 and to 3.74 K for $x$ = 0.25. Compared to the absolute shift of $T_N$ in samples with Co and Ir concentration up to $x$ = 0.25 [10, 16, 17], one can immediately see that the effect of Pd substitution is weaker. The Sommerfeld coefficient obtained from the $C/T = \gamma + \beta T^2$ fit in the interval 10 K < $T$ < 15 K yields $\gamma \sim 350$ mJ mole$^{-1}$ K$^{-2}$ for CeRhIn$_5$ which is somewhat lower than the $\gamma$ value of 400 mJ mole$^{-1}$ K$^{-2}$ reported in Ref. [19]. With increasing Pd concentration, $\gamma$ decreases to about 270 mJ mole$^{-1}$ K$^{-2}$ for the sample with $x$ = 0.25. The phonon term $\beta T^2$ was subtracted from the raw $C/T$ data to allow an estimate of the magnetic entropy. Below $T_N$, this procedure gives for CeRhIn$_5$ a 4$f$ contribution of about 0.27 $R$ ln2 to the entropy $S_{4f}$ and, for the $x = 0.25$ sample, an entropy $S_{4f} \sim$ 29% $R$ ln2, which is probably caused by Kondo-screened ordered moments as was suggested previously [19], providing similar values for CeRhIn$_5$. The remaining entropy of more than 70% $R$ ln2 is recovered at $T \sim$ 13 K, as presented in the inset of figure 2.

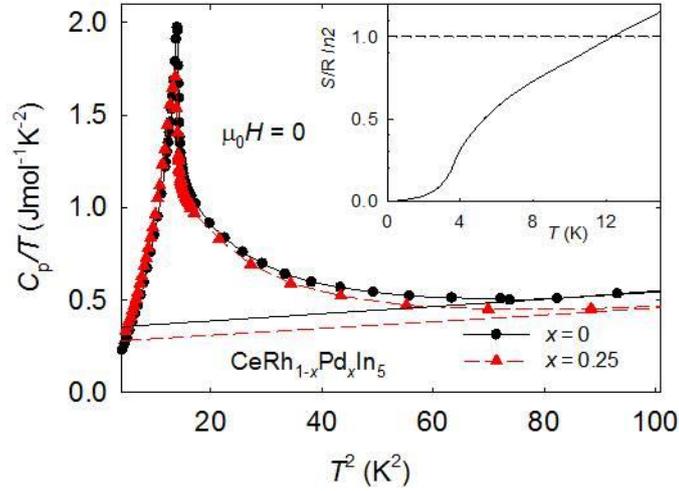

**Figure 2.** Total specific heat of CeRh$_{1-x}$Pd$_x$In$_5$ for $x = 0$ and 0.25 in zero magnetic field. The lines represent the $C/T = \gamma + \beta T^2$ fits. The inset shows the temperature evolution of the magnetic entropy of CeRhIn$_5$.

The effect of the applied magnetic field is strongly anisotropic as can be seen in figures 3(*b*) and 3(*c*). The specific heat in a magnetic field of 9 T applied along the *a*- and *c*-axis is presented in detail for CeRh$_{1-x}$Pd$_x$In$_5$ ($x = 0.1$ and 0.25) in comparison with CeRhIn$_5$.

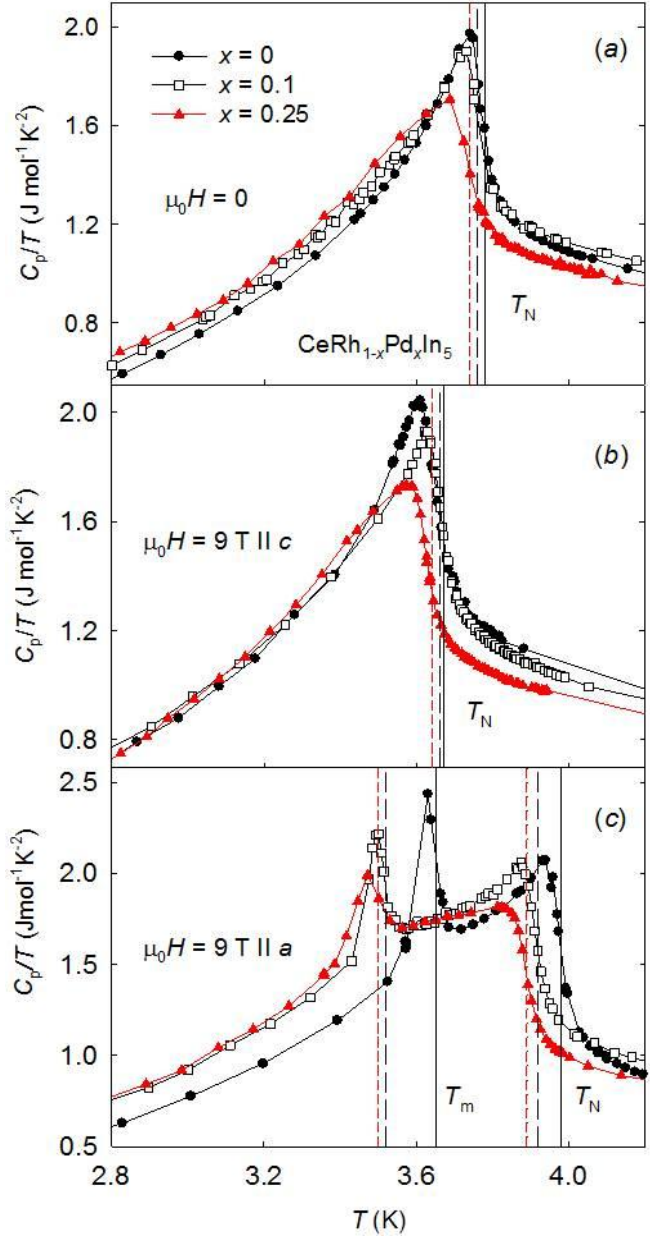

**Figure 3.** Low-temperature region of the specific heat of CeRh$_{1-x}$Pd$_x$In$_5$ for $x = 0$, 0.1 and 0.25 in zero and 9 T magnetic field applied along both crystallographic directions. The vertical lines indicate the evolution of the Néel temperature $T_N$ and the magnetic-field-induced transition at $T_m$ as a function of Pd content and magnetic field.

The Néel temperature $T_N$ slightly decreases in Pd-substituted samples while no other magnetic transition appears with increasing magnetic field applied along the *c*-axis. On the other hand, $T_N$ is increased ($T_N = 3.87$ K at 9 T) for CeRh$_{0.75}$Pd$_{0.25}$In$_5$ when the magnetic field is applied within the basal plane. Moreover, another two magnetic-field-induced transitions at $T_1 = 2.68$ K and $T_2 = 3.43$ K appear at 2.6 T and 2.5 T, respectively. To trace the evolution of these transitions in the basal plane thoroughly, we measured the low-temperature specific heat of CeRh$_{0.75}$Pd$_{0.25}$In$_5$ in small magnetic-field steps in the range from 2.5 T to 3.5 T as shown in figure 4. Although the induced transitions are not pronounced as clearly as in CeRhIn$_5$ [18], likely due to the disorder caused by Pd substitution, they can be traced up to 3 T mimicking the behavior of CeRhIn$_5$. Above 3 T, except the transition at $T_N$, only one transition remains at $T_m$, increasing gradually up to $T_m = 3.47$ K at 9 T.

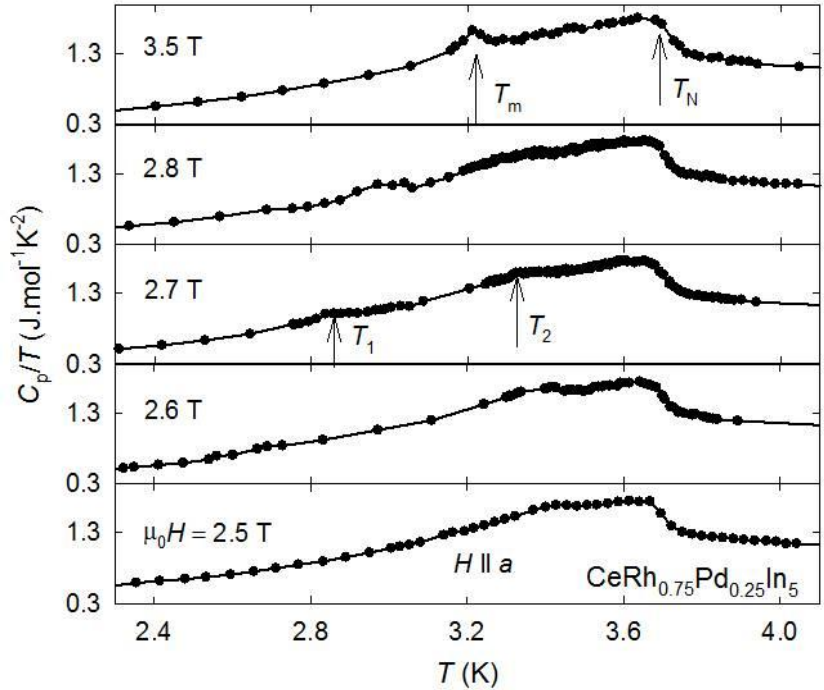

**Figure 4.** Low-temperature specific heat of CeRh$_{0.75}$Pd$_{0.25}$In$_5$ measured in various magnetic fields applied along the *a*-axis. The arrows point to the transition temperatures $T_N$, $T_1$, $T_2$ and $T_m$.

Having results available from the specific heat measurements, the *H-T* magnetic phase diagram for CeRhIn$_5$ and CeRh$_{0.75}$Pd$_{0.25}$In$_5$ can be constructed for *H* // *a* (see figure 5). The close resemblance of the magnetic phase diagrams of CeRhIn$_5$ and CeRh$_{0.75}$Pd$_{0.25}$In$_5$ suggests analogous character of the phases I, II and III (same notation as in Ref. [4]). For CeRhIn$_5$, the regions I and II present incommensurate helicoidal AF structures with the Ce magnetic moment spiraling transversely along the *c*-axis (region II is characterized by a larger magnetic moment), while region III corresponds to a commensurate magnetic structure with the magnetic moment parallel to the *c*-axis [4]. To the best of our knowledge, the effect of the magnetic field has hardly been studied for substituted CeRh$_{1-x}$T$_x$In$_5$ compounds with CeRh$_{0.9}$Ir$_{0.1}$In$_5$ as the only exception [10]. The phase diagram for CeRh$_{0.75}$Pd$_{0.25}$In$_5$ resembles qualitatively the one with 10% Ir substitution, although the latter one is shifted to higher temperatures. The remarkable similarity of both phase diagrams leads to the natural conclusion that substitution at the transition-metal site does not affect the magnetic structure for moderate Pd or Ir [10] concentration. To sum up the results obtained from specific heat, the Pd substitution does not significantly influence the magnetic behavior reflecting probably only slight variations of the RKKY interaction. This scenario is similar to the one presented in case of the Co and Ir substitutions [7, 8, 17].

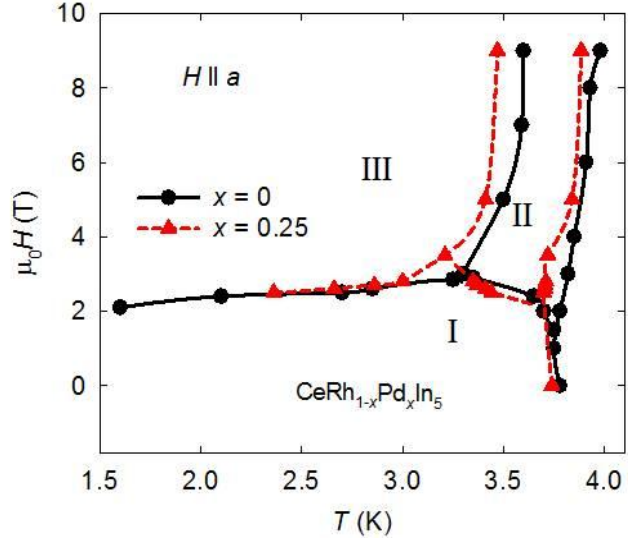

**Figure 5.** $H$-$T$ diagram for CeRhIn$_5$ and CeRh$_{0.75}$Pd$_{0.25}$In$_5$ for the magnetic field applied along the $a$-axis. The lines are guides to the eyes, representing the evolution of the Néel temperature and the magnetic-field-induced transitions. For CeRhIn$_5$, the regions marked I, II refer to an incommensurate AF structure while, in region III, the AF structure is commensurate with the lattice. The same notation is used as in Raymond *et al.* [4].

The temperature dependence of the susceptibility of CeRh$_{0.75}$Pd$_{0.25}$In$_5$ was measured in magnetic fields of 0.5 T and 3 T applied along the $a$- and $c$-axis (figures 6($a$) and 6($b$), respectively). All curves show a subtle change of slope at $T_N$ indicating the transition from the paramagnetic to the AF phase. In addition, the $a$-axis susceptibility shows a significant dip at $T_2 = 3.3$ K which reflects most likely the transition between phases I and II (see figure 4). Well above $T_N$, the susceptibility exhibits a maximum at about 7.5 K, independent of the direction of the magnetic field. No significant shift of the susceptibility maximum is observed compared to the data reported for CeRhIn$_5$ [19]. The origin of the susceptibility enhancement at low temperatures is ascribed to the presence of degenerate $J = 5/2$ Kondo impurities [27].

In order to study the influence of the crystal field and the consequent magnetocrystalline anisotropy, the magnetic susceptibility $\chi(T)$ of CeRh$_{0.75}$Pd$_{0.25}$In$_5$ was measured up to room temperature along both principal directions (see inset of figure 6($a$)). At temperatures above 100 K (200 K) the $a$-axis ($c$-axis) susceptibility, $\chi(T)$ obeys the Curie-Weiss law with an effective moment of $\mu_{eff} = 2.2\ \mu_B$/Ce (2.3 $\mu_B$/Ce). The experimental $\mu_{eff}$ values are somewhat smaller than the Hund's rule value of 2.54 $\mu_B$ for a free Ce$^{3+}$ ion. The reduced $\mu_{eff}$ values together with the pronounced deviation of $1/\chi(T)$ from linear behavior at lower temperatures, can be attributed to a strong crystal-field interaction, similar to what has been reported for CeRhIn$_5$ [19]. The paramagnetic Curie temperatures derived from the $a$- and $c$-axis susceptibility are $\theta_P^a = -52$ K and $\theta_P^c = 41$ K. The difference $|\theta_P^c - \theta_P^a|$ (table 2) demonstrates a significant anisotropy of the magnetic susceptibility which seems to be little affected by the Pd substitution (see comparison of $|\theta_P^c - \theta_P^a|$ for CeRh$_{0.75}$Pd$_{0.25}$In$_5$ and CeRhIn$_5$).

**Table 2.** Concentration dependence of the paramagnetic Curie temperature $\theta_P$ and the effective magnetic moment $\mu_{eff}$ of $CeRh_{1-x}Pd_xIn_5$. The results for $x = 0$ are taken from Ref. [19].

|  | $\theta_P$ (K) | | $\mu_{eff}$ ($\mu_B$/Ce) | |
| --- | --- | --- | --- | --- |
| $x$ | $\parallel a$ | $\parallel c$ | $\parallel a$ | $\parallel c$ |
| 0 | -79 | 16 | 2.4 | 2.4 |
| 0.25 | -52 | 41 | 2.2 | 2.3 |

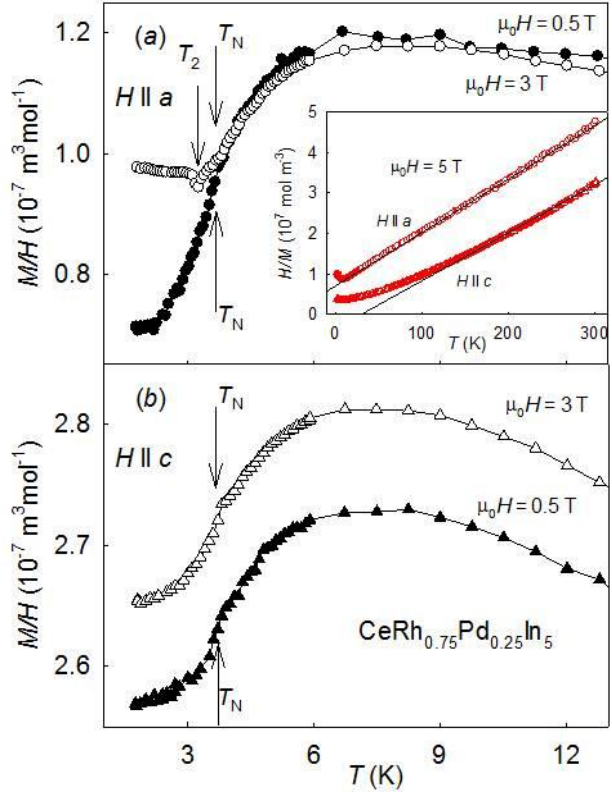

**Figure 6.** Temperature dependence of $M/H$ of $CeRh_{0.75}Pd_{0.25}In_5$ measured in magnetic fields of 0.5 T and 3 T applied along the $a$-axis ($a$) and $c$-axis ($b$). The arrows indicate the transitions at $T_N$ and $T_2$. The values of the transition temperatures have been taken from specific-heat data. Inset: Temperature dependence of $H/M$ of $CeRh_{0.75}Pd_{0.25}In_5$ measured in a magnetic field of 5 T oriented along the $a$- and $c$-axis, respectively (black lines represent the Curie-Weiss fits).

To investigate the effect of hydrostatic pressure, the electrical resistivity of the sample with $x = 0.25$ has been measured in pressures up to 2.25 GPa. The temperature dependences of the electrical resistivity of $CeRh_{0.75}Pd_{0.25}In_5$ at various pressures are shown in figure 7($a$). Similar to $CeRhIn_5$, the temperature dependence of resistivity of the $CeRh_{0.75}Pd_{0.25}In_5$ single crystal exhibits a maximum at $T_{max}$ for hydrostatic pressures larger than 1 GPa. The temperature $T_{max}$ initially decreases, reaching its minimum value at ~ 1.4 GPa and then increases with further increasing pressure (see the $T$-$p$ phase diagram in figure 8). The minimum on the $T_{max}$ vs. $p$ plot coincides with the decay of AF ordering and the gradual emergence of superconductivity, which points to an onset of Kondo coherence, observed also in the case of $CeRhIn_5$ [19].

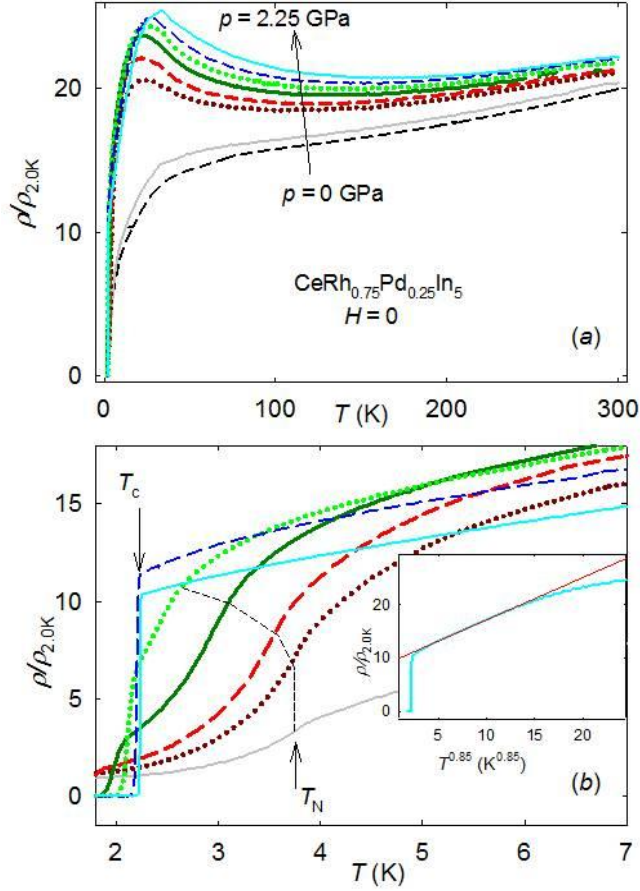

**Figure 7.** (*a*) The temperature dependence of the electrical resistivity of CeRh$_{0.75}$Pd$_{0.25}$In$_5$ for current along the *a*-axis at hydrostatic pressures of 0, 0.34, 1.24, 1.42, 1.66, 1.82, 1.99 and 2.25 GPa. (*b*) Low-temperature part of the electrical resistivity of CeRh$_{0.75}$Pd$_{0.25}$In$_5$. The arrows indicate the superconducting temperature $T_c$ given by the onset of superconductivity at $p = 1.99$ GPa and the Néel temperature $T_N$ at $p = 0.34$ GPa. The dashed line shows the pressure dependence of $T_N$, determined from the inflection point of the resistivity-temperature curve. The estimate of $T_c$ at $p = 1.42$ GPa is based on onset-like behavior since there are no data below 1.8 K. The inset presents a linear fit of the resistivity at 2.25 GPa in a $\rho$ vs. $T^{0.85}$ plot which is consistent with nFl behavior.

At the first sight, one can recognize similarities of the evolution of magnetism and superconductivity with applied pressure for CeRh$_{0.75}$Pd$_{0.25}$In$_5$ and CeRhIn$_5$ which imply similar physics (see figure 7(*b*)): a) the antiferromagnetism remains nearly intact at pressures up to about 1.3 GPa ($T_N$ remains nearly constant up to 1.24 GPa), b) superconductivity emerges above 1.3 GPa where $T_N$ starts to decrease, c) above $T_c$, at $p = 2.25$ GPa, the resistivity exhibits a $T^n$-dependence with $n = 0.85$ over a temperature range from about 4.5 to 11 K, resembling the nFl behavior in CeRhIn$_5$ [28]. The SC temperature $T_c$ initially increases with increasing pressure, whereas $T_N$ decays, pointing to a correlation between antiferromagnetism and superconductivity, similar to the case of CeRhIn$_5$ [21]. The SC transition is clearly seen at $T_c = 1.97 \pm 0.10$ K at a pressure of 1.66 GPa (although the low-temperature part of the resistivity curve at 1.42 GPa indicates that superconductivity with $T_c <$ 1.8 K may emerge already at this pressure). The values of $T_N$ and $T_c$ become equal at the critical pressure $p_{c1}$ of about 2 GPa, which is close to the value reported in Ref. [21]. Higher pressures (> 2 GPa) lead to suppression of the AF order and to a further slight increase of the SC temperature to $T_c = 2.23 \pm 0.02$ K at $p = 2.25$ GPa. This value corresponds to $T_c = 2.22 \pm 0.02$ K at $p = 2.35$ GPa for CeRhIn$_5$ [29]. Extrapolation of the pressure dependence of $T_N$ of CeRhIn$_5$ yields $T_N = 0$ at $p_{c2} \sim 2.6$ GPa where probably only the SC phase survives. In analogy with CeRhIn$_5$, we tentatively draw a similar conclusion for CeRh$_{0.75}$Pd$_{0.25}$In$_5$.

From the $T$-$p$ phase diagram in figure 8, it is clear, that the effect of Pd substitution up to $x = 0.25$ in CeRhIn$_5$ on the magnetic behavior is almost negligible which is somewhat unusual compared to the Ir and Co substitution [7, 17] on the $T$-metal site. From the bulk modulus and the change of the unit cell volume the change in pressure of about -0.5 GPa can be estimated leading to a shift of the phase diagram by this pressure difference. In case of Pd substitution, the reason is probably given by the compensation effect of the electronic structure and the crystal lattice.

However, inspecting the evolution of $T_c$ with pressure, one immediately recognizes a striking difference which is more expectable according to previous paragraph. In particular, the SC temperature of CeRh$_{0.75}$Pd$_{0.25}$In$_5$ at 1.66 GPa is almost three times higher than the corresponding value observed in the specific heat of CeRhIn$_5$ [29]. Although various experimental methods (specific heat, resistivity, NQR) may lead to slightly different values of $T_c$, the difference observed for CeRh$_{0.75}$Pd$_{0.25}$In$_5$ and CeRhIn$_5$ is far outside these variations. The difference is even more striking because the enhanced $T_c$ value is observed in the substituted compound where one would expect that atomic disorder of the transition-metal (Rh, Pd) sublattice should rather lead to suppression of a cooperative phenomenon like superconductivity. On the other hand, we should note that the highest measured value of $T_c$ of CeRh$_{0.75}$Pd$_{0.25}$In$_5$ where the superconducting temperatures start to saturate ($T_c = 2.24$ K at $p \sim 2.25$ GPa) and the maximum $T_c$ value of CeRhIn$_5$ ($T_c = 2.22$ K at $p \sim 2.35$ GPa), are very similar and both at approximately the same pressure value. According to the work of Park et al. [22], different values of $T_c$ obtained from thermodynamic and transport measurements point to presence of an intermediate SC region which is textured in real space due to a coexisting magnetic ordering. This textured SC phase is intercalated with magnetic domains below the critical pressure $p_{c1}$ causing significant anisotropy in the in-plane and out-of-plane resistivity. Although the resistivity of CeRh$_{0.75}$Pd$_{0.25}$In$_5$ could be measured only for the in-plane orientation of electrical current due to the dimensions of the sample, the data allow us to speculate that the electron doping by Pd prevent the electron structure to form the filamentary superconductivity proposed in CeRhIn$_5$.

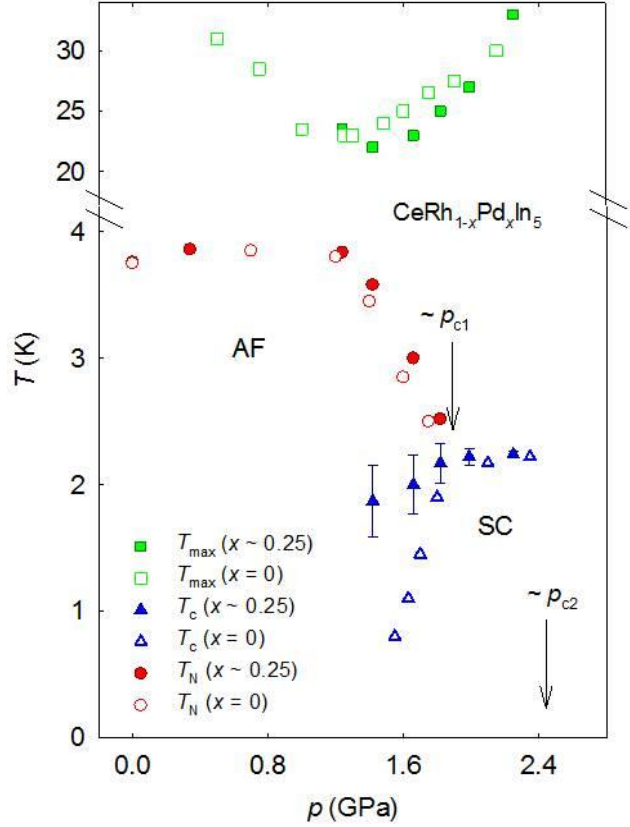

**Figure 8.** $T$-$p$ phase diagram for $CeRh_{0.75}Pd_{0.25}In_5$ and $CeRhIn_5$. $T_{max}$ is the temperature where the resistivity reaches its maximum. The values of $T_{max}$, $T_N$ and $T_c$ for $CeRhIn_5$ have been taken from Ref. [21, 29].

## 4. Conclusions

$CeRh_{1-x}Pd_xIn_5$ single crystals (for $x$ up to 0.25) have been grown by means of the In self-flux method in order to study effects of Pd substitution for Rh on magnetism and superconductivity compared to behavior of $CeRhIn_5$. The tetragonal $HoCoGa_5$-type of structure was found conserved ($c/a$ ratio is intact) with the Pd substitution. Data from measurements of magnetic, transport and thermodynamic properties at ambient pressure complemented by low-temperature electrical resistivity study in hydrostatic pressure can be concluded as follows. Similar to the results observed for the Co and Ir substitution, also the Pd substitution does not significantly influence the magnetic behavior. The Néel temperature $T_N$ as well as the magnetic-field-induced transitions $T_1$, $T_2$ and $T_m$ are only slightly shifted to lower temperatures with increasing the Pd concentration and evolution of these transitions with magnetic fields is analogous to results reported for $CeRhIn_5$ [18]. Therefore we tentatively conclude the magnetic structure in the substituted systems to be similar to $CeRhIn_5$. The pressure evolution of $T_N$ resembles strongly that for $CeRhIn_5$ as well [21].

On the other hand, the pressure dependence of the superconducting temperature is noticeably influenced by the introduction of Pd although the highest measured value of $T_c$ above 2 GPa is very similar to the maximum $T_c$ value of SC dome in $CeRhIn_5$. This observation indicates that the $CeRh_{1-x}Pd_xIn_5$ system becomes superconducting at lower pressures leading to an extended region of coexistence of AF and SC phase. Thus, the present work suggests that, at higher Pd concentrations, the superconducting state may already exist at ambient pressure, similarly to the CeRh$_{1-x}$Co$_x$In$_5$ and CeRh$_{1-x}$Ir$_x$In$_5$ systems. Compared to the filamentary SC phase which was discussed in CeRhIn$_5$ [22], the Pd substitution might not prefer the emergence of textured superconducting planes. The possible occurrence of an ambient superconducting phase in the *T-x* phase diagram of CeRh$_{1-x}$Pd$_x$In$_5$ and the anisotropy in electrical resistivity deserves further investigation which means development of techniques suitable for single crystal preparation and transport measurement on samples of submilimeter size.

**Acknowledgements**

The authors thank Frank de Boer for fruitful discussion and critical remarks. This work was supported by the Grant Agency of Charles University (Project no. 134214) and by the Czech Science Foundation (Project no. GP14-17102P). Experiments were performed at Magnetism and Low Temperatures Laboratories MLTL (http://mltl.eu/), which are supported within the program of Czech Research Infrastructures (Project no. LM2011025).

# References

[1] Thompson J D, Movshovich R, Fisk Z, Bouquet F, Curro N J, Fisher R A, Hammel P C, Hegger H, Hundley M F, Jaime M, Pagliuso P G, Petrovic C, Phillips N E and Sarrao J L 2001 J. Magn. Magn. Mater. **226-230** 5

[2] Tursina A, Nesterenko S, Seropegin Y, Noel H and Kaczorowski D 2013 J. Sol. State Chem. **200** 7

[3] Kratochvilova M, Dusek M, Uhlirova K, Rudajevova A, Prokleska J, Vondrackova B, Custers J and Sechovsky V 2014 J. Cryst. Growth **397** 47

[4] Raymond S, Ressouche E, Knebel G, Aoki D, and Flouquet J 2007 J. Phys.: Condens. Matter **19** 242204

[5] Bao W, Pagliuso P G, Sarrao J L, Thompson J D, Fisk Z, Lynn J W and Erwin R W 2000 Phys. Rev. B **62** R14621

[6] arXiv:1408.6585v1

[7] Nicklas M, Sidorov V A, Borges H A, Pagliuso P G, Sarrao J L and Thompson J D 2004 Phys. Rev. B **70** 020505(R)

[8] Pagliuso P G, Movshovich R, Bianchi A D, Nicklas M, Moreno N O, Thompson J D, Hundley M F, Sarrao J L and Fisk Z 2002 Physica B **312** 129

[9] Mendonca Ferreira L, Bittar E M, Pagliuso P G, Hering E N, Ramos S M, Borges H A, Baggio-Saitovich E, Bauer E D, Thompson J D and Sarrao J L 2007 Physica C **460–462** 672–673

[10] Light B E, Kumar R S, Cornelius A L, Pagliuso P G and Sarrao J L 2004 Phys. Rev. B **69** 024419

[11] Pham L D, Park T, Maquilon S, Thompson J D and Fisk Z 2006 Phys. Rev. Lett. **97** 056404

[12] Booth C H, Bauer E D, Bianchi A D, Ronning F, Thompson J D, Sarrao J L, Cho J Y, Chan J Y, Capan C and Fisk Z 2009 Phys. Rev. B **79** 144519

[13] Bauer E D, Ronning F, Maquilon S, Pham L. D, Thompson J D and Fisk Z 2008 Physica B **403** 1135–1137

[14] Bauer E D, Mixson D, Ronning F, Hur N, Movshovich R, Thompson J D, Sarrao J L, Hundley M F, Tobash P H and Bobev S 2006 Physica B **378-380** 142

[15] Raymond S, Buhot J, Ressouche E, Bourdarot F, Knebel G and Lapertot G 2014 Phys. Rev. B **90** 014423

[16] Pagliuso P G, Petrovic C, Movshovich R, Hall D, Hundley M F, Sarrao J L, Thompson J D and Fisk Z 2001 Phys. Rev. B **64** 100503

[17] Zapf V S, Freeman E J, Bauer E D, Petricka J, Sirvent C, Frederick N A, Dickey R P and Maple M B 2001 Phys. Rev. B **65** 014506

[18] Cornelius A L, Pagliuso P G, Hundley M F and Sarrao J L 2001 Phys. Rev. B **64** 144411

[19] Hegger H, Petrovic C, Moshopoulou E G, Hundley M F, Sarrao J L, Fisk Z and Thompson J D 2000 Phys. Rev. Lett. **84** 4986

[20] Mito T, Kawasaki S, Kawasaki Y, Zheng G Q, Kitaoka Y, Aoki D, Haga Y and Ōnuki Y 2003 Phys. Rev. Lett. **90** 077004

[21] Knebel G, Aoki D, Brison J P, Howald L, Lapertot G, Panarin J, Raymond S and Flouquet J 2010 phys. stat. sol. (b) **247** 557

[22] Park T, Lee H, Martin I, Lu X, Sidorov V A, Gofryk K, Ronning F, Bauer E D and Thompson J D 2012 Phys. Rev. Lett. **108** 077003

[23] Uhlirova K, Prokleska J, Sechovsky V and Danis S 2010 Intermetallics **18** 2025

[24] Fujiwara N, Matsumoto T, Koyama-Nakazawa K, Hisada A and Uwatoko Y 2007 Rev. Sci. Instrum. **78** 073905

[25] Murata K, Yoshino H, Yadav H O, Honda Y and Shirakawa N 1997 Rev. Sci. Instrum. **68** 2490

[26] Ou M N, Gofryk K, Baumbach R E, Stoyko S S, Thompson J D, Lawrence J M, Bauer E D, Ronning F, Mar A and Chen Y Y 2013 Phys. Rev. B **88** 19

[27] Rajan V T 1983 Phys. Rev. Lett. **51** 308

[28] Park T, Sidorov V A, Ronning F, Zhu J-X, Tokiwa Y, Lee H, Bauer E D, Movshovich R, Sarrao J L and Thompson J D 2008 Nature **456** 366

[29] Park T, Tokiwa Y, Ronning F, Lee H, Bauer E D, Movshovich R and Thompson J D 2010 phys. stat. sol. (b) **247** 553